\documentclass[twocolumn]{article} 
\usepackage{times}  
\usepackage{helvet}  
\usepackage{courier}  
\usepackage[hyphens]{url}  
\usepackage{graphicx} 
\usepackage{natbib}  
\usepackage{caption} 
\usepackage{algorithm}
\usepackage{algorithmic}
\usepackage{authblk}
\usepackage{url}

\usepackage{newfloat}
\usepackage{listings}
\DeclareCaptionStyle{ruled}{labelfont=normalfont,labelsep=colon,strut=off} 
\floatstyle{ruled}
\newfloat{listing}{tb}{lst}{}
\floatname{listing}{Listing}

\DeclareCaptionStyle{ruled}{labelfont=normalfont,labelsep=colon,strut=off} 
\floatstyle{ruled}
\newfloat{listing}{tb}{lst}{}
\floatname{listing}{Listing}

\usepackage{subfigmat}

\usepackage{booktabs}
\usepackage{url}

\usepackage{xcolor}

\title{Avatar Communication Provides More Efficient Online Social Support Than Text Communication}

\author[1,6]{Masanori Takano}
\author[2]{Kenji Yokotani}
\author[3]{Takahiro Kato}
\author[4]{Nobuhito Abe}
\author[5]{Fumiaki Taka}

\affil[1]{CyberAgent, Inc.}
\affil[2]{Tokushima University}
\affil[3]{Kyushu University}
\affil[4]{Kyoto University}
\affil[5]{Toyo University}
\affil[6]{Keio University}

\usepackage{bibentry}

\begin{document}

\maketitle

\begin{abstract}
Online communication via avatars provides a richer online social experience than text communication. This reinforces the importance of online social support. Online social support is effective for people who lack social resources because of the anonymity of online communities. We aimed to understand online social support via avatars and their social relationships to provide better social support to avatar users. Therefore, we administered a questionnaire to three avatar communication service users (Second Life, ZEPETO, and Pigg Party) and three text communication service users (Facebook, X, and Instagram) ($N=8,947$). There was no duplication of users for each service. By comparing avatar and text communication users, we examined the amount of online social support, stability of online relationships, and the relationships between online social support and offline social resources (e.g., offline social support). We observed that avatar communication service users received more online social support, had more stable relationships, and had fewer offline social resources than text communication service users. However, the positive association between online and offline social support for avatar communication users was more substantial than for text communication users. These findings highlight the significance of realistic online communication experiences through avatars, including nonverbal and real-time interactions with co-presence. The findings also highlighted avatar communication service users' problems in the physical world, such as the lack of offline social resources. This study suggests that enhancing online social support through avatars can address these issues. This could help resolve social resource problems, both online and offline in future metaverse societies.
\end{abstract}

\section{Introduction}

Social support improves mental health~\cite{Cohen1985,Rothon2011}.
This is also true online (online social support)~\cite{Mesch2006,Chung2013,ChoudhuryDe2014,Trepte2015,Cole2017,Sharma2018,Reblin2018,MasanoriTakano2019,Yokotani2021a,Pierce2020,takano_icwsm2022}.
Online social support is particularly practical for people who lack social resources because of the anonymity of online communities~\cite{Kang2016,ChoudhuryDe2014}.
Detailed disclosure facilitates the reception of greater social support in cases of negative experiences and emotions~\cite{ChoudhuryDe2014}.
The characteristics of online communication facilitate disclosure ~\cite{Kang2016,ChoudhuryDe2014} by decreasing listeners’ fears of rejection ~\cite{Mesch2006,Kuster2015,Andalibi2016}.
This is helpful that people face issues that are difficult to disclose in the physical world owing to stigma and prejudice (sexual minorities~\cite{Yokotani2021a}, bullying victimization~\cite{Cole2017,MasanoriTakano2019,takano_icwsm2022}, sexual abuse~\cite{Andalibi2016}, suicidal feelings~\cite{ChoudhuryDe2014}, and mental health problems~\cite{Sharma2018}).

Nonverbal communication is essential for social support because socioemotional cues play a crucial role in emotional support and are conveyed nonverbally~\cite{Mehrabian1970,Manusov2006}.
Previous studies~\cite{Ledbetter2008,Trepte2015,McCloskey2015} have highlighted the lack of nonverbal communication in online text communication.
Interactions using online tools are less satisfactory than face-to-face interactions~\cite{Kock2005,Vlahovic2012}.
Although people have invented emotional expressions under the constraints of text media, such as emoticons and emojis, the effects of online social support are limited~\cite{Ledbetter2008,Trepte2015,McCloskey2015}.

By contrast, avatar communication, in which individuals with virtual bodies can display facial expressions and gestures in a virtual space, enables nonverbal and real-time interactions with online co-presence~\cite{Antonijevic2008,Green-Hamann2011,OConnor2015,vanderLand2011ModelingEnvironments}.
Avatar actions realize emotional expressions when talking about bullying in the physical world, such as the painfulness of bullying and empathy for painfulness, in the avatar communication application Pigg Party~\cite{MasanoriTakano2019}.
In the metaverse, users' social presence tends to increase because of the avatar’s spatial presence~\cite{vanderLand2011ModelingEnvironments}.
\citet{Oh2023SocialLoneliness} indicated that a high social presence increases online social support, which mitigates loneliness in two metaverse platforms (ZEPETO and Roblox).
Proxemic behavior in the online virtual world game Second Life has been observed to evoke feelings similar to those experienced in the physical world~\cite{Antonijevic2008} and to facilitate communication in social support groups (alcoholics anonymous and cancer caregivers)~\cite{Green-Hamann2011}.
These avatar communication features provide improved social support~\cite{Green-Hamann2011,takano_icwsm2022}.

Therefore, online social support through avatars is necessary for those who lack social resources in the physical world.
Online social support will play an essential role in future metaverse societies.
We aimed to understand online social support via avatars and their social relationships to provide better social support to avatar users.
We compared three avatar communication services (Second Life, ZEPPETO, and Pigg Party) with three text communication services (Facebook, X, and Instagram).

While there are several previous studies for online social support, most of them focus on a single application (E-mail: \cite{Ledbetter2008}, Reddit: \cite{ChoudhuryDe2014,Andalibi2016,Sharma2018}, Facebook: \cite{McCloskey2015}, World of Warcraft: \cite{OConnor2015}, Second Life: \cite{Antonijevic2008,Green-Hamann2011}, Pigg Party: \cite{MasanoriTakano2019,Yokotani2021a,takano_icwsm2022}) or compare text communication applications with each other (Facebook, Instagram, Snapchat, and Twitter~\cite{Phua2017}) or avatar communication applications with each other (ZEPETO and Roblox~\cite{Oh2023SocialLoneliness}).
As far as the authors know, no studies compare multiple avatar and text communication applications on the same basis. 
Although there have been studies examining the effects of text and avatars on self-disclosure in controlled environments using applications developed specifically for psychological experiments~\cite{Yuan2025}, there is a lack of research investigating self-disclosure and social support on applications used in everyday life.
In this context, two significant gaps need to be addressed:  
First, how do social relationships differ quantitatively between text-based communication applications and avatar-based communication applications?
Second, despite their unique specifications, design, and user types, are there common characteristics shared within every kind of application?
For these research gaps, the comparison between the two types of applications can provide essential insights into the perspective of generalizing the findings from previous research.

First, we examine the following hypotheses:
\begin{itemize}
    \item {\bf H1}: Online social support is practical in avatar communication services compared with text communication services.
\end{itemize}

Second, we considered the stability of the relationships in each communication service.
Real-time interaction with online co-presence in avatar communication~\cite{Antonijevic2008,Green-Hamann2011,OConnor2015} restricts the time and online location compared with asynchronous text communication because avatar communication requires users to be in the same place simultaneously for communication.
This indicates that they have stable social relationships.
Therefore, we propose the following hypothesis:
\begin{itemize}
    \item {\bf H2}: The users of avatar communication services have stable relationships compared with text communication services.
\end{itemize}
Examining this hypothesis is important for understanding online social support features via avatars because stable relationships are essential for social support.
Stable and close relationships provide rich social support~\cite{Westmyer2009,Kendrick2012,Roberts2014}, and several relationships offer opportunities to receive social support.
In addition, online social support can be reinforced in virtual worlds by improving online ego network structures~\cite{Perez-Aldana2021,takano_icwsm2022}.
People tend to disclose their emotions and sensitive information in a closed cyberspace with friends (not strangers)~\cite{Jaidka2018,MasanoriTakano2019}.

Third, we focused on the relationship between online and offline social support.
Positive correlations have been frequently observed between online and social support in the physical world (offline social support)~\cite{Trepte2015,Lin2018,takano_icwsm2022}. 
This finding is significant because it suggests that online social support enhances social support in the physical world~\cite{McKenna1998,Lin2018,Thomas2020}.
This is because receiving online social support facilitates offline social activities, thereby increasing the number of relationships that provide offline social support~\cite{Lin2018}.
Consequently, high levels of social support in the virtual world can mitigate loneliness by mediating social support in the physical world~\cite{Thomas2020} and might decrease bullying victimization.
Therefore, we propose the following hypothesis:
\begin{itemize}
    \item {\bf H3}: The users of avatar communication services have offline social resources (high offline social support, low loneliness, and less likely to be bullied) if H1 is true.
\end{itemize}

The contributions of this study are as follows:
\begin{itemize}
\item This is the first study comparing multiple avatar and text communication applications on the same basis.
\item Avatar communication service users received more online social support than text communication service users (H1 was supported).
\item They maintained more stable social relationships than text communication service users (H2 was supported).
\item They had fewer offline social resources than text communication service users (H3 was not supported), i.e., their social resources tended to be lacking.
\item Their association between online and offline social resources was higher than that of text communication users. This suggests that improving their online social support may solve the problem of a lack of offline social resources.
\end{itemize}
We expect to contribute to the resolution of social resource problems in future metaverse societies.

\section{Methods}
\subsection{Participants}
Participants were recruited from a panel managed by Cross Marketing, Inc., consisting of individuals living in Japan. The survey was conducted in Japan and administered in Japanese, so most participants were presumed to be native Japanese speakers residing in Japan.
Thus, the impact of cultural differences arising from the regions of the application providers (for example, Zepeto in South Korea and Pigg Party in Japan) on the analysis results seems small.
First, they answered questions about the services they used and their frequency of use from the three avatar communication services and three social media/social networking services. 
If they used any of the services at least once a month, they were assigned to one of them and answered the items described later.
There was no duplication of users for each service.
Participants were equally divided in terms of age and gender.
Table~\ref{tbl_demog} summarizes the basic statistics of each service.

\begin{table}[t!]
  \begin{center}
    \small
\begin{tabular}{ll|rrr}											\toprule
Service	&	Gender	&	N	&	Mean	&	Std. Dev. \\ \midrule
Second Life	&	Female	&	522	&	38.56	&	11.30 \\ 
	&	Male	&	960	&	45.21	&	12.15 \\ 
	&	Other	&	13	&	34.46	&	13.57 \\ \midrule
ZEPETO	&	Female	&	584	&	38.98	&	12.38 \\ 
	&	Male	&	890	&	44.78	&	11.45 \\ 
	&	Other	&	17	&	37.88	&	12.82 \\ \midrule
Pigg Party	&	Female	&	605	&	38.87	&	11.72 \\ 
	&	Male	&	869	&	45.66	&	12.21 \\ 
	&	Other	&	19	&	42.68	&	12.82 \\ \midrule
Facebook	&	Female	&	619	&	46.82	&	12.69 \\ 
	&	Male	&	859	&	48.35	&	12.53 \\ 
	&	Other	&	3	&	28.00	&	7.21 \\ \midrule
X	&	Female	&	697	&	39.43	&	13.36 \\ 
	&	Male	&	787	&	40.19	&	12.17 \\ 
	&	Other	&	7	&	30.71	&	6.52 \\ \midrule
Instagram	&	Female	&	1,017	&	38.45	&	13.26 \\ 
	&	Male	&	472	&	40.34	&	13.64 \\ 
	&	Other	&	7	&	41.14	&	15.91 \\ \midrule
Total & - &  8,947 & 42.20 & 12.90 \\ \bottomrule
 \end{tabular}
  \caption{Gender and age of each service participant}
    \label{tbl_demog}
\end{center}
\end{table}

\subsection{Communication Services}
\subsubsection{Avatar Communication Services}

\begin{figure}[t!]
\vspace{3mm}
  \begin{center}
      \begin{subfigmatrix}{1}
          \subfigure[Second Life]{\includegraphics[width=0.9\columnwidth]{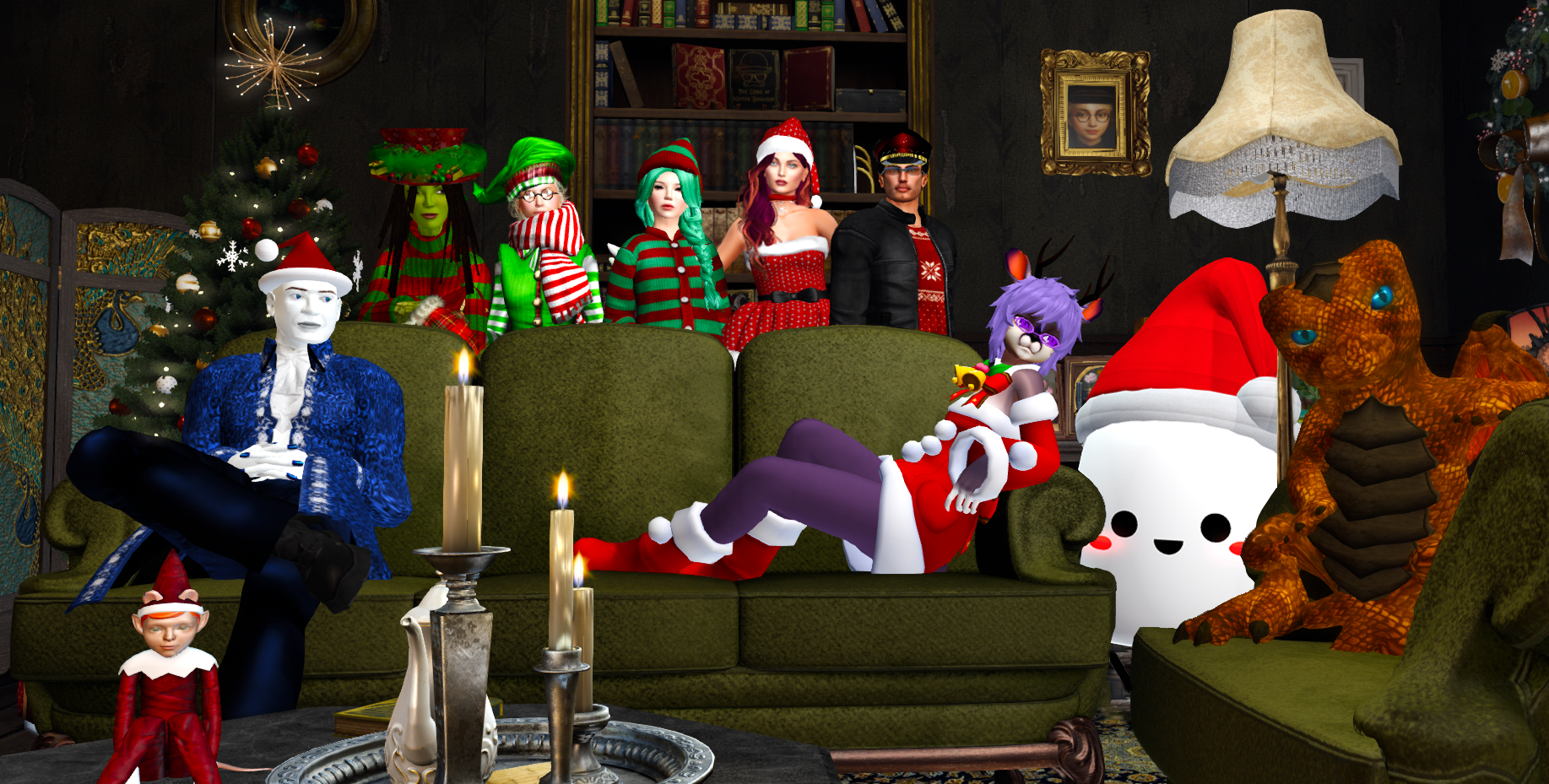}}
          \subfigure[ZEPETO]{\includegraphics[width=0.9\columnwidth]{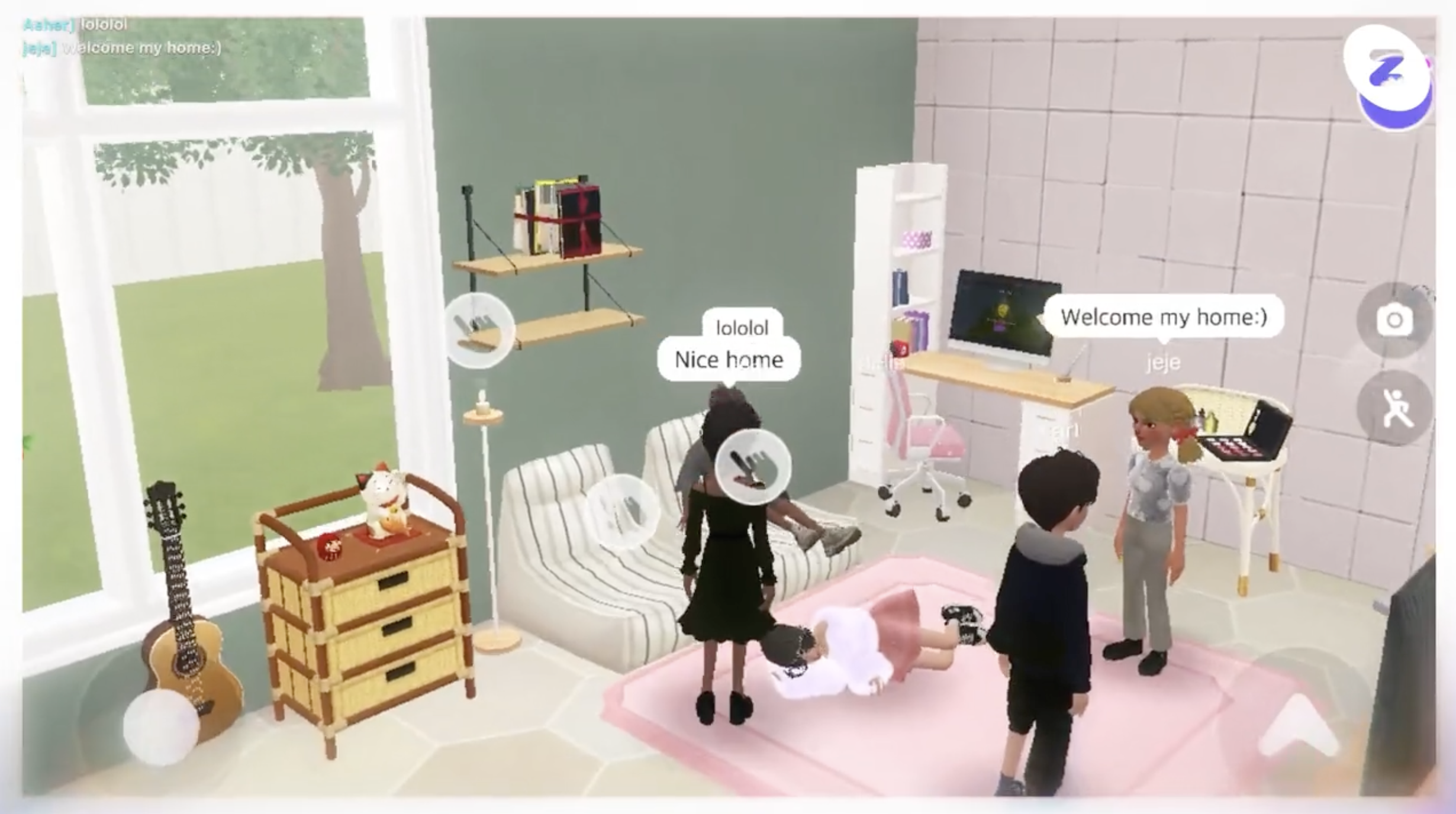}}
          \subfigure[Pigg Party]{\includegraphics[width=0.9\columnwidth]{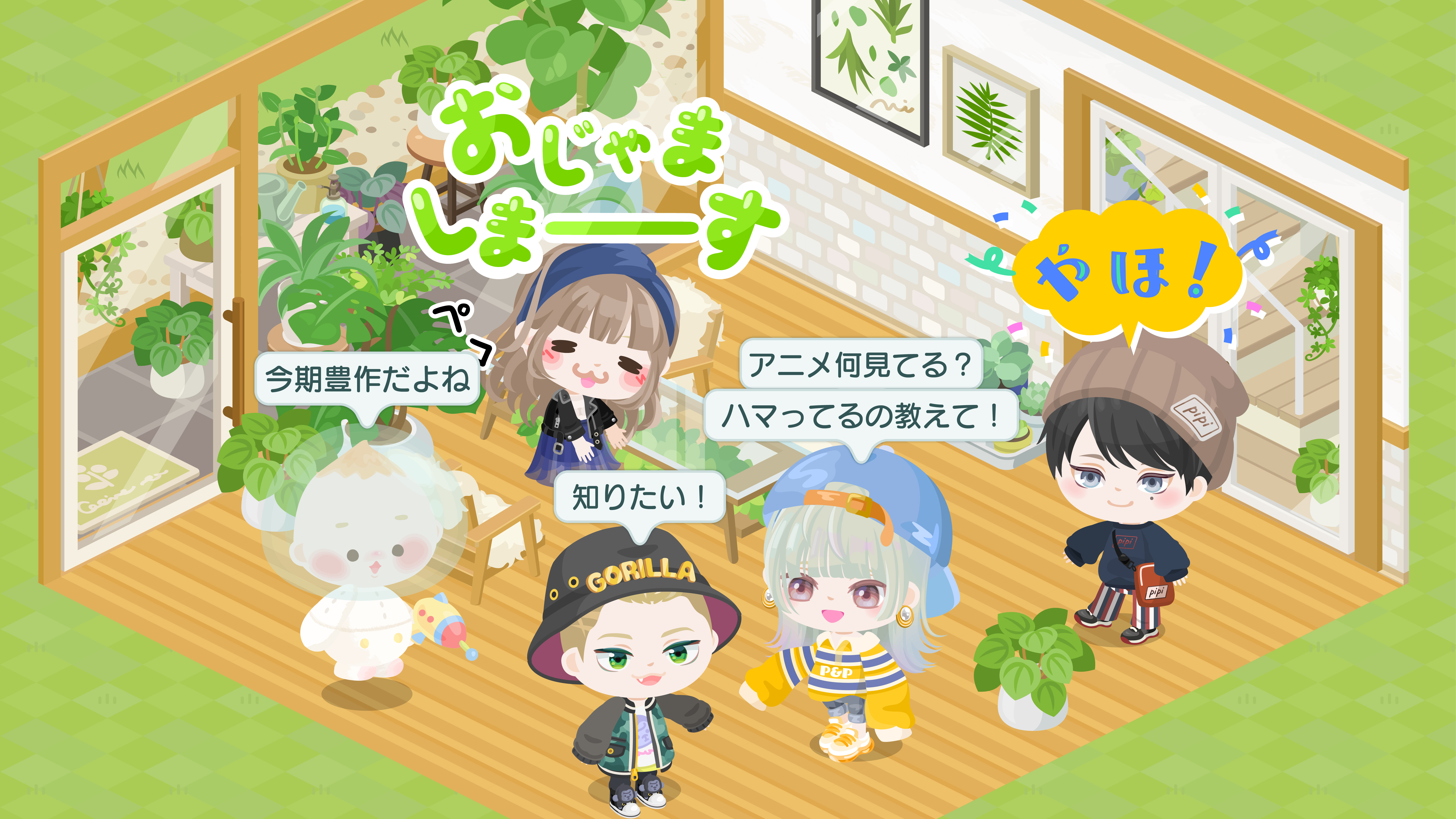}}
      \end{subfigmatrix}
      \caption{Screen shots of avatar communication services (image sources: Second Life: \url{https://community.secondlife.com/blogs/entry/14709-season's-greetings-from-linden-lab/}, ZEPETO: \url{https://x.com/zepeto_official/status/1617327573234589698}, and Pigg Party:~\citet{Yokotani2021_chb2}). }   \label{fig_shots}
  \end{center}
\end{figure}

We surveyed users of three metaverse/avatar communication services: Second Life, ZEPETO, and Pigg Party (Fig.~\ref{fig_shots}).
This allows them to create avatars, explore virtual worlds, and interact with others through their avatars using text and voice chats.
Second Life provides realistic 3D avatars, Pigg Party provides fancy 2D avatars, and ZEPETO provides an intermediate between the two avatars.
Several researchers have studied human behavior in the metaverse and gained various insights based on data from Second Life: \citet{Antonijevic2008,Green-Hamann2011,Greiner2014,Mitra2016,Messinger2019}, ZEPETO: \citet{Lee2023,Aris2023,Ko2024,Hur2024,Lee2024b}, and Pigg Party: \citet{MasanoriTakano2019,Yokotani2021a,takano_icwsm2022,Yokotani2022cp,Yokotani2024}.

Second Life, operated by Linden Research, Inc. (US), has provided metaverse services since 2003. 
It has approximately 900,000 active monthly users\footnote{\url{https://nwn.blogs.com/nwn/2020/06/sl-mau-pandemic-june-2020.html}}.
In 2009, 1.3\% of its users were Japanese\footnote{\url{https://www.itmedia.co.jp/news/articles/0703/07/news074.html}}.

ZEPETO, operated by Naver Z (South Korea), is a metaverse application launched in 2018. 
It boasts of 20 million monthly active users\footnote{\url{https://www.kedglobal.com/metaverse/newsView/ked202203040009}}, 70\% of whom are females aged from 10 to early 20s. Japanese users are estimated to comprise 5--10\% of the total user base\footnote{\url{https://xtrend.nikkei.com/atcl/contents/casestudy/00012/00921/}}.

Pigg Party is a social avatar community app operated by CyberAgent, Inc. (Japan), which was launched in 2015.
A previous study reported on at least 550,000 active players over six months~\cite{Yokotani2021_chb2}.
Pigg Party players were predominantly young women, with a female ratio of 61\% and a teenage ratio of 65\%~\cite{MasanoriTakano2019}.

\subsubsection{Text Communication Services}
We conducted a survey with X, Facebook, and Instagram users as major social media and social networking services.
These services provide three types of communication methods (sending direct messages, posting one’s feed, and replying to posts).
Each service has hundreds of millions of users. 
Several researchers have studied online human behavior in these services.

\subsection{Measure}

\subsubsection{Perceived Online/Offline Social Support}
We examined two types of social support (emotional and instrumental)~\cite{Fukuoka1997} and three sources of social support (family, offline friends, and online friends) because these types and sources influence the buffering effects of social support on mental health~\cite{Malecki2003,Semmer2008,Rothon2011,McCloskey2015,oriol2017,Wills2015,Trepte2015,Liu2018}.
We categorized family and offline friends as sources of offline social support and online friends as sources of online social support.

We classified social support into emotional and instrumental support~\cite{Declercq2007,Shakespeare-Finch2011,Tsuboi2016,Obst2019}.
This classification is the most comprehensive despite the availability of various definitions of social support~\cite{Shakespeare-Finch2011,Obst2019}.

In addition to online friends as sources of social support, we considered the relationships between recipients and sources, including family and offline friends.
These relationships are broadly applicable across different contexts, unlike teachers and classmates, who are specific to student surveys.
We selected these types because the participants had diverse backgrounds.

We performed a confirmatory factor analysis using maximum likelihood estimation for this perceived social support scale.
Cronbach's $\alpha$ values for emotional and instrumental social support from family were $0.940$ and $0.937$, those from offline friends were $0.940$ and $0.921$, and those from online friends were $0.949$ and $0.959$, respectively.

In this study, we used the results of a principal component analysis (PCA) for perceived emotional and instrumental social support for each source type according to a previous study~\cite{takano_icwsm2022}.
This approach was selected because emotional and instrumental support are often highly correlated ~\cite{Semmer2008,Shakespeare-Finch2011,Tsuboi2016,Obst2019,takano_icwsm2022}, which can lead to multicollinearity in mental health regression analyses ~\cite{Tsuboi2016}.
The correlations between the two types of social support were $0.939$ (family), $0.889$ (offline friends), and $0.772$ (online friends).

The PCA identified the same two principal components of social support: family, offline friends, and online friends.
The first principal component represented the overall strength of perceived social support by integrating the two types of social support.
For all social support sources, the coefficient of this component was $0.707$.
The second principal component represented the relative strength of perceived emotional support compared to perceived instrumental support.
For all social support sources, the coefficient of this component was $-0.707$.
The first principal component was used to represent the strength of perceived social support.

\subsubsection{Online Relational Mobility}
\citet{Yuki2007} developed a scale to measure relational mobility or the general number of opportunities to form new relationships, when necessary, in a given society or social context. 
They identify two factors in meeting new people and selecting interaction partners for relational mobility.
We used meetings with new people on each online communication service to measure online relational mobility.
This would be negatively correlated with the stability of social relationships.
We performed a confirmatory factor analysis using the maximum likelihood estimation of this scale. 
Cronbach's $\alpha$ value for online relational mobility was $0.846$.

\subsubsection{Loneliness}
We used the Japanese three-item loneliness scale (TIL scale)~\cite{Igarashi2019}, which is based on the TIL scale~\cite{Hughes2004}.
We performed a confirmatory factor analysis using the maximum likelihood estimation of this scale. 
The Cronbach's $\alpha$ value for loneliness was $0.831$.

\subsubsection{Bullied Experience in the Physical World}

We used a Japanese bully/victim questionnaire~\cite{Yokotani2021a} based on the Revised Olweus Bully/Victim Questionnaire~\cite{Olweus1996} to evaluate offline victimization.
We performed a confirmatory factor analysis using the maximum likelihood estimation of this scale. 
The Cronbach's $\alpha$ value for offline victimization was $0.960$.

\subsubsection{Control Variables}

The participants provided their demographic information (gender (female, male, and others) and age (18 years or older) and usage frequency of services (six levels: almost every day, 4--5 days per week, 2--3 days per week, a day per week, 2--3 days per month, and a day per month).
These were used as control variables.

\subsection{Ethics Approval Statement}

This study was approved by the ethics committee of CyberAgent, Inc (CAE-2023-07). 
All procedures were performed in accordance with the guidelines for studies involving human participants and the ethical standards of the Institutional Research Committee. 

Participants provided informed consent to participate in the survey and could stop at any time (refer to Fig.~\ref{fig_ss_informed_consent}). Participants were also allowed to withdraw their responses after completing the survey.
This informed consent form included an inquiry contact form for requests for disclosure and withdrawal of responses.

Quantitative data outputs are presented at the aggregate level, indicating that no identifying information is presented.

\section{Results}

We revealed the features of the social relationships of avatar communication services compared with text communication services.

\subsection{Perceived Online Social Support (H1)}

To examine H1, we compared perceived online social support between avatar and text communication service users. 
Fig.~\ref{fig_oss} shows the results of variance (ANOVA) for perceived online social support with controlling age, gender, and usage frequency, where Facebook was used as the reference category (that is, $Coef.=0$ shows Facebook).
All avatar communication service users displayed more online social support than text communication users.
Refer to Table~\ref{tbl_oss} in the appendix for the statistical tests.
We also obtained the following ANOVA results in the same settings:

\begin{figure}[t!]
  \begin{center}
    \includegraphics[width=0.95\columnwidth]{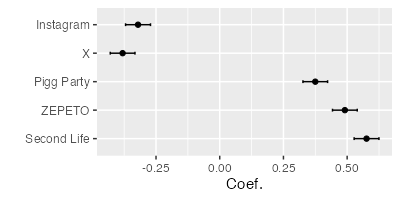}
    \caption{
    Result of ANOVA for perceived online social support, where $F(5, 8937) = 130.899; p\mathrm{-value} < 0.001$.
    Points show coefficients and error bars show standard errors, with Facebook as a reference category.
      The following figures also indicate the same presentation of this figure.
    }
    \label{fig_oss}
  \end{center}
\end{figure}

\subsection{Online Relational Mobility (H2)}

\begin{figure}[t!]
  \begin{center}
    \includegraphics[width=0.95\columnwidth]{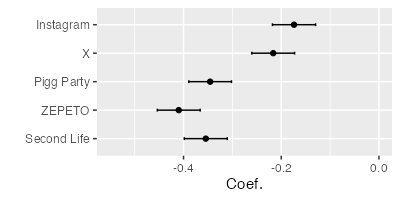}
    \caption{
        Result of ANOVA for online relational mobility, where $F(5, 8937) = 43.81; p\mathrm{-value} < 0.001$
}
    \label{fig_mob}
  \end{center}
\end{figure}

To examine H2, we compared online relational mobility between avatar and text communication service users. Fig.~\ref{fig_mob} shows the results of ANOVA for online relational mobility while controlling for age, gender, and usage frequency.
All avatar communication service users exhibited lower online relational mobility than text communication service users; specifically, they had more stable relationships than text communication service users.
See Table~\ref{tbl_mob} in the appendix for the statistical tests.

\subsection{Offline Social Support, Loneliness, and Bullying Victimization (H3)}

\begin{figure}[t!]
  \begin{center}
    \includegraphics[width=0.95\columnwidth]{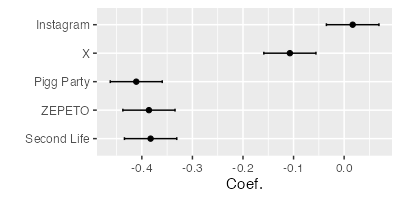}
    \caption{
        Result of ANOVA for perceived online social support from family, where $F(5, 8937) = 42.631; p\mathrm{-value} < 0.001$ and $F(5, 8937) = 16.391; p\mathrm{-value} < 0.001$, respectively}
    \label{fig_ss_fam}
  \end{center}
\end{figure}

\begin{figure}[t!]
  \begin{center}
    \includegraphics[width=0.95\columnwidth]{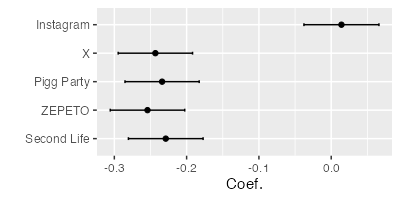}
    \caption{
        Result of ANOVA for perceived online social support from offline friends, where $F(5, 8937) = 42.631; p\mathrm{-value} < 0.001$ and $F(5, 8937) = 16.391; p\mathrm{-value} < 0.001$, respectively}
    \label{fig_ss_fri}
  \end{center}
\end{figure}

\begin{figure}[t!]
  \begin{center}
    \includegraphics[width=0.95\columnwidth]{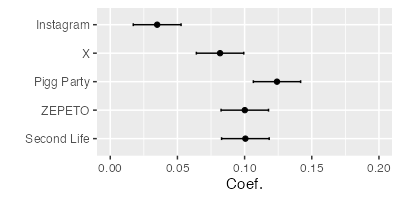}
    \caption{
            Result of ANOVA for loneliness, where $F(5, 8937) =  19.299; p\mathrm{-value} < 0.001$
    }
    \label{fig_loneliness}
  \end{center}
\end{figure}

\begin{figure}[t!]
  \begin{center}
    \includegraphics[width=0.95\columnwidth]{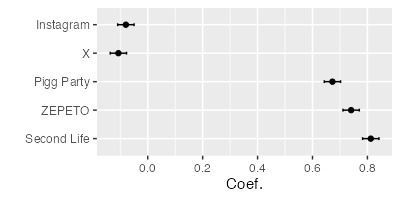}
    \caption{        Result of ANOVA for bullying victimization, where $F(5, 8937) = 465.833; p\mathrm{-value} < 0.001$}
    \label{fig_vic}
  \end{center}
\end{figure}

To examine H3, we compared perceived social support from social relationships in the physical world (family and offline friends), loneliness, and bullying victimization between avatars and text communication service users. 
Figs.~\ref{fig_ss_fam}, \ref{fig_ss_fri}, \ref{fig_loneliness}, and \ref{fig_vic} show the result of ANOVA for perceived social support from family and offline friends, loneliness, and bullying victimization, respectively, after controlling for age, gender, and usage frequency.
Refer to Tables~\ref{tbl_ss_fam}, \ref{tbl_ss_fri}, \ref{tbl_Loneliness}, and \ref{tbl_bullying} in the appendixes for statistical tests.
Avatar communication service users displayed less social support from family and offline friends, more loneliness, and more victimization than text communication users, excluding X user’s social support from offline friends and X users’ loneliness.
In contrast to H3, the social resources for avatar communication users are lacking.
Note that X users were also lacking compared with Facebook and Instagram users.

\subsection{Association Between Online Social Support and Offline Social Support}

The above analysis suggests a negative association between online and offline social support inter-services.
To mitigate the lack of offline social resources, we analyzed the association between online and offline social support intra-services.

\begin{figure}[t!]
  \begin{center}
    \includegraphics[width=0.95\columnwidth]{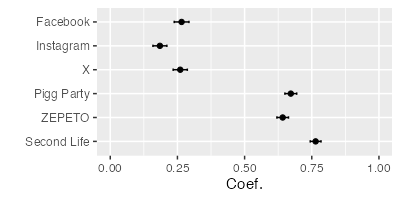}
    \caption{
    Association between online social support and social support from family
    }
    \label{fig_coef_fam_ss}
  \end{center}
\end{figure}

\begin{figure}[t!]
  \begin{center}
    \includegraphics[width=0.95\columnwidth]{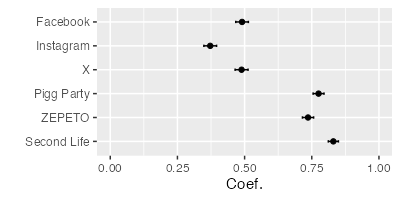}
    \caption{Association between online social support and social support from offline friends
    }
    \label{fig_coef_fri_ss}
  \end{center}
\end{figure}

Figs.~\ref{fig_coef_fam_ss} and \ref{fig_coef_fri_ss} show the coefficient of regression for each service, where independent variables were social support from family and offline friends, and the dependent variable was online social support, with controlling age, gender, and usage frequency.
Refer to Tables~\ref{tbl_oss_famss} and \ref{tbl_oss_friss} in the Appendix for the statistical tests.

All the services exhibited positive associations between online and offline social support.
Avatar communication services exhibit stronger associations than text communication services.


\section{Discussion}

We examined three hypotheses to understand the characteristics of online social support via avatars by comparing them with text communication services.

Avatar communication users perceived online social support more than text communication services.
H1 was supported.
This suggests that the similarity of communication via avatars to face-to-face communication (nonverbal expressions, real-time interaction, and co-presence~\cite{Antonijevic2008,Green-Hamann2011,OConnor2015}) is essential for online social support and that social relationships in avatar communication services are suited to online social support.
This is because these characteristics contribute to conveying emotions and facilitate social support~\cite{Antonijevic2008,Green-Hamann2011,MasanoriTakano2019}.
Although online communication is inferior in terms of emotional interaction and social support~\cite{Kock2005,Ledbetter2008,Vlahovic2012,Trepte2015,McCloskey2015}, avatar communication may eliminate this issue.

Avatars can enhance online social support~\cite{Green-Hamann2011,takano_icwsm2022}.
\citet{MasanoriTakano2019} analyzed bullying consultation conversations on Pigg Party, highlighting online social support through avatars. 
The study found that bullying victims often used avatar actions to express emotions and self-disclose. Recipients also used emotional expressions through text and avatars. For instance, victims used words like ``distress,'' ``cutting-off,'' and ``suicidal feelings,'' along with avatar actions like ``wailing'' to show their suffering. Listeners responded with similar words and avatar actions to show empathy.

Avatar communication users had more stable relationships than text communication service users; that is, H2 was supported.
This seems to be because the characteristics of avatar communication (real-time interaction with online co-presence~\cite{Antonijevic2008,Green-Hamann2011,OConnor2015}) require simultaneous communication.
Such stable relationships can be fundamental to constructing a society where people have close relationships~\cite{Roberts2014}.
However, text communication services tend to develop weak bonds that help bridge online communities~\cite{Ahmad2023}.
Therefore, the formation of social networks and social interactions may differ between avatars and text communications.

We found these differences in social relationships between avatar and text communication applications despite each application group having different specifications, avatar appearances, and user demographics.
This suggests that social relationships created on the applications are qualitatively different due to the kinds of communication, i.e., via avatars or texts.
Comparison between two groups with different communication styles on the same basis enabled us this finding.

Online social support facilitates offline social resources.
However, avatar communication service users who received more online social support than text communication service users tended to lack social resources.
Their perceived offline social support was lower, they felt lonely, and they tended to experience bullying victimization.
Thus, H3 was not supported.
This may be because there are people using avatar communication services as escape places~\cite{Yokotani2021a,Lee2023,Hur2024}.
These findings suggest that avatar users may require additional offline social resources and highlight differences in user segments and motivations for use (e.g., seeking information or alleviating loneliness). 
Future longitudinal studies are required for the separation of such predisposition from the effects of the application usage.

Note that X users also tend to lack offline social resources (low offline social resources and high loneliness).
This may be because Japanese X users tend to think of socializing as bothersome compared with Facebook and Instagram users~\cite{watanabe2019}.

However, the association between online and offline social support was positive in the intraservice context, which is consistent with the findings of previous studies~\cite{Trepte2015,Lin2018,takano_icwsm2022}.
This association was stronger for avatar communication services than for text communication services.
X, which tended to lack offline social resources similar to avatar communication services, did not exhibit such a strong association.
Previous studies have shown that receiving online social support can enhance offline social support~\cite{McKenna1998,Liu2013,Thomas2020}, suggesting that avatar communication services can enhance offline social resources for those who lack offline social resources.
Actually, social relationship construction via avatars in Pigg Party reduces social anxiety of sexual and gender minorities in the physical world~\cite{yokotani_ajsp}.

These findings were shared among three avatar communication services offering different avatar creatives, such as realistic or fancy and 3D or 2D.
In other words, the fundamental effects of avatar communication were unrelated to avatar creativity, as revealed by our research findings.
It will also be valuable to indicate the relationship between avatar creativity and the psychological effects on communication by comparing several types of avatars.

On the other hand, text communication applications also have advantages in online social support. 
Since they are asynchronous communication, people can post their self-disclosure and responses whenever they want, and it is easier for both the self-discloser and the supporter to communicate after careful consideration. 
For instance, on Reddit, there are subreddits where people with depression feelings can write, and their stories are heard, effectively providing online social support~\cite{ChoudhuryDe2014}. 
Additionally, social media platforms like X, Facebook, and Instagram are suitable for bridging relationships~\cite{Phua2017,Takano2018} (weak ties~\cite{Granovetter1973}). 
Therefore, while avatar communities excel in providing emotional support such as empathy and respect, text media has advantages in terms of obtaining new information and perspectives. 
For example, individuals whose parents did not graduate from college can interact with college graduates on Facebook, gain information about universities, and improve their chances of attending college~\cite{Wohn2013}.
These differences might generate variations in the motivations for using each application.

What can platforms do to enhance the offline resources of avatar communication service users?
Improving ego networks for increasing sharing friends, that is, belonging to a densely-connected community, can increase social support~\cite{KIM1991,Roberts2014}.
This is also true of online social support~\cite{takano_icwsm2022,Hygen2024}.
It seems crucial for platforms to facilitate communities in virtual worlds and help users feel that they belong to them.

Three approaches can be considered to improve ego networks:
The first involves friend recommendations to increase sharing among friends.
Several online communication platforms, including avatar communication services, provide potential friend recommendations~\cite{Tang2013}.
The friend-of-friend algorithm, which increases the number of friends shared between users, is a major approach for friend recommendations.

The second aspect is the sharing of places of communication.
The co-presence of avatar communication requires sharing of places among users.
These shared places facilitate the formation of partnerships within a community~\cite{Godinho2015,Bhatt2020} by creating densely connected communities~\cite{TakanoJPC2021}.
Therefore, it would be effective for the platform to create a place where users wish to gather and a mechanism for them to become attached to that place.

Third, social interactions can be indirectly facilitated through avatar interventions.
Avatar customization increases avatar identification~\cite{Birk2016,Mancini2018,takano_chbr_2022} and high avatar identification enhances sociality and belonging to a community~\cite{Vasalou2007,VanReijmersdal2013,Kao2018,takano_chbr_2022}.
In particular, avatar customization, as in real life (e.g., mimicking a friend’s clothes and changing hairstyle), reinforces avatar identification~\cite{takano_chbr_2022}.
Users who design avatars to reflect their desired attractive selves have become sociable~\cite{Messinger2008,Messinger2019}.
Therefore, it would be effective for platforms to hold campaigns that facilitate avatar customization, such as in real life and making wishful avatars.
For instance, the Halloween campaign can allow users to dress up their avatars and interact with online friends, similar to how they would in the physical world. This would enhance avatar identification in the virtual world and likely increase communication.

We also expect these approaches, which facilitate communication~\cite{Birk2016,Birk2019,takano_chbr_2022}, to improve the platforms' key performance indicators because they enhance user engagement.
In other words, the facilitation of online social support by platforms benefits not only the users but also the platforms themselves. Platforms can conduct this through as friend recommendations or seasonal campaigns, which approaches are extensions of the existing functionalities of the application.
This implies that such promotion of online social support can be a seamless and acceptable approach for both platforms and their users.

Furthermore, the existence of users who used avatar communication services as escape places suggests that avatar communication services can provide psychological safety places for people lacking offline social resources.
Actually, sexual and gender minorities who tend to lack offline social resources acquire social resources in Pigg Party~\cite{Yokotani2021a}.
Facilitating avatar communication services for those lacking offline social resources can improve their online social resources and mental health.

Finally, we summarize our theoretical contributions as follows.
First, we indicated the importance of online nonverbal communications for online social support with outer validity, although previous studies~\cite{Antonijevic2008,Green-Hamann2011,OConnor2015,MasanoriTakano2019,takano_icwsm2022} investigated this topic in one platform.
This study contributes to the psychology of nonverbal communication.

Second, the nature of online social networks may differ significantly between avatars and text communication platforms. 
This difference affects the quality of online social support.
Online communication provides communication partners with impressions different from those in offline communication~\cite{Kock2005,Vlahovic2012,Burke2016}.
People use these methods depending on their communication purposes and partners~\cite{Burke2014}.
Consequently, qualitative differences emerged between online and offline social networks ~\cite{Takano2018}.
This may be because, in the hierarchy of social relationships (the Dunbar circle~\cite{Zhou2005,Dunbar2018}), the level of social relationships that an individual can make depends on the means of communication.
Online communication via avatars in the virtual world may have forms similar to offline communication, such as online text communication.
This will contribute to social network theory and social psychology regarding friendships.

Third, we found that the relationship between online and offline social support depends on avatar or text communication.
Previous studies~\cite{McKenna1998,Liu2013,Thomas2020} have indicated that online social support can enhance offline social support.
We demonstrated that online social support via avatars had stronger associations with offline social support than with text.
This study contributes to the field of social support psychology.

\subsection{Limitations and Future Works}

We obtained our findings by examining and comparing three avatars and three text communication services.
The mechanism of these findings must be analyzed using detailed user behavioral log data.
Additionally, qualitative research, such as interviews, can provide deeper insights into online social support via avatars.

This study investigated three avatar communication applications and three text communication applications. 
There are many ways to communicate online, e.g., avatars in virtual reality and online bulletin boards. 
Recent advancements in AI technology have made it possible to create avatars that look very similar to the user and synchronize facial expressions and gestures. 
By conducting similar studies on these diverse communication methods and ultra-realistic avatars, such as Vasa-1, Live Portrait, or Hallo, we can better understand the scope of our findings. 

The participants in this study were Japanese speakers living in Japan, which limits cultural diversity. 
Verifying our findings across multiple cultures would help clarify their applicability.

We studied the associations between online social support, online relational stability, and offline social resources in avatars and text communication services.
An experimental study to improve online social support by manipulating social relationships, virtual communication locations, and avatar customization should be conducted to determine the intervention outcomes for ego networks.

High avatar identification enables users to enjoy games and communication in the virtual world~\cite{VanReijmersdal2013,Birk2016,Kao2018}.
However, it has also been shown to increase the risk of game addiction~\cite{Mancini2018}.
Therefore, we have to clarify not only how to facilitate online social resources but also the relationship between this and addiction risks.

Our study builds on previous research~\cite {McKenna1998, Lin2018, Thomas2020} and aims to shed light on the effects of online social support on offline social support, thereby potentially deepening our understanding of this relationship.
On the other hand, we can also consider that individuals with high sociality receive high levels of online and offline social support.
Examining the causality of this phenomenon is also required in avatar communication.

This study deals with communication between humans through avatars. 
One possible approach to further enhance this is the introduction of AI-powered communication agents. 
For example, an AI agent could participate in human conversations to facilitate smooth dialogue and suppress aggressive remarks, thereby maintaining a psychologically safe environment and enhancing online social support between humans.

\section{Conclusion}

To understand online social support through avatars, we examined three hypotheses regarding online social support, online relational mobility, and offline social resources based on comparisons between avatars and text communication services.
Our findings indicate the importance of realistic online communication experiences (nonverbal and real-time interactions with co-presence) and avatar communication service users' problems in the physical world (lack of offline social resources).
In addition, it is suggested that these problems can be resolved by enhancing online social support through avatars.
This could contribute to online and offline social resource problems in future metaverse societies.

\section*{Acknowledgement}
This work was supported by JST, PRESTO Grant Number JPMJPR2367, Japan.

\bibliographystyle{unsrt}

\appendix
\clearpage
\setcounter{equation}{0}
\setcounter{section}{0}
\setcounter{figure}{0}
\setcounter{table}{0}
\renewcommand{\thesection}{A\arabic{section}}

\renewcommand{\figurename}{Fig.}
\renewcommand{\thefigure}{A\arabic{figure}}
\renewcommand{\tablename}{Table.}
\renewcommand{\thetable}{A\arabic{table}}

\renewcommand{\theequation}{A\arabic{equation}}

\section*{Appendixes}

\begin{table*}[h!]
  \begin{center}
    \small
\begin{tabular}{ll|rrrl}											\toprule
Avatar	&	Text	&	Coef.	&	Std. Err.	&	p-value	&		\\	\midrule
Second Life	&	Facebook	&	0.5761	&	0.0487	&	$<$0.001	&	***	\\	
	&	X	&	0.9572	&	0.0502	&	$<$0.001	&	***	\\	
	&	Instagram	&	0.8967	&	0.0508	&	$<$0.001	&	***	\\	\midrule
ZEPETO	&	Facebook	&	0.4910	&	0.0486	&	$<$0.001	&	***	\\	
	&	X	&	0.8721	&	0.0499	&	$<$0.001	&	***	\\	
	&	Instagram	&	0.8116	&	0.0503	&	$<$0.001	&	***	\\	\midrule
Pigg Party	&	Facebook	&	0.3749	&	0.0484	&	$<$0.001	&	***	\\	
	&	X	&	0.7560	&	0.0497	&	$<$0.001	&	***	\\	
	&	Instagram	&	0.6955	&	0.0501	&	$<$0.001	&	***	\\	\bottomrule
\end{tabular}
  \caption{Result of Tukey's range test in analysis of variance (ANOVA) for perceived online social support with controlling age, gender, and usage frequency, where $F(5, 8937) = 130.899; p\mathrm{-value} < 0.001$.
  Positive coefficients indicate that avatar communication services exhibit larger coefficients than text communication services.
  The following tables also indicate the same presentation as this table.
  }
    \label{tbl_oss}
\end{center}
\end{table*}

\begin{table*}[h!]
  \begin{center}
    \small
\begin{tabular}{ll|rrrl}											\toprule
Avatar	&	Text	&	Coef.	&	Std. Err.	&	p-value	&		\\	\midrule
Second Life	&	Facebook	&	-0.3545	&	0.0440	&	$<$0.001	&	***	\\	
	&	X	&	-0.1380	&	0.0454	&	0.0285	&	*	\\	
	&	Instagram	&	-0.1806	&	0.0459	&	0.0012	&	**	\\	\midrule
ZEPETO	&	Facebook	&	-0.4098	&	0.0439	&	$<$0.001	&	***	\\	
	&	X	&	-0.1934	&	0.0451	&	$<$0.001	&	***	\\	
	&	Instagram	&	-0.2360	&	0.0454	&	$<$0.001	&	***	\\	\midrule
Pigg Party	&	Facebook	&	-0.3456	&	0.0437	&	$<$0.001	&	***	\\	
	&	X	&	-0.1291	&	0.0449	&	0.0467	&	*	\\	
	&	Instagram	&	-0.1717	&	0.0453	&	0.0021	&	**	\\	\bottomrule
\end{tabular}
  \caption{
  Result of Tukey's range test in ANOVA for online relational mobility with controlling age, gender, and usage frequency, where $F(5, 8937) = 43.81; p\mathrm{-value} < 0.001$.
}
    \label{tbl_mob}
\end{center}
\end{table*}

\begin{table*}[h!]
  \begin{center}
    \small
\begin{tabular}{ll|rrrl}											\toprule
Avatar	&	Text	&	Coef.	&	Std. Err.	&	p-value	&		\\	\midrule
Second Life	&	Facebook	&	-0.3827	&	0.0517	&	$<$0.001	&	***	\\	
	&	X	&	-0.2754	&	0.0534	&	$<$0.001	&	***	\\	
	&	Instagram	&	-0.3996	&	0.0540	&	$<$0.001	&	***	\\	\midrule
ZEPETO	&	Facebook	&	-0.3860	&	0.0516	&	$<$0.001	&	***	\\	
	&	X	&	-0.2787	&	0.0530	&	$<$0.001	&	***	\\	
	&	Instagram	&	-0.4029	&	0.0535	&	$<$0.001	&	***	\\	\midrule
Pigg Party	&	Facebook	&	-0.4111	&	0.0514	&	$<$0.001	&	***	\\	
	&	X	&	-0.3038	&	0.0529	&	$<$0.001	&	***	\\	
	&	Instagram	&	-0.4280	&	0.0532	&	$<$0.001	&	***	\\	\bottomrule
\end{tabular}
  \caption{
  Result of Tukey's range test in ANOVA for Perceived Social Support from Family with controlling age, gender, and usage frequency, where $F(5, 8937) = 42.631; p\mathrm{-value} < 0.001$.
  }
    \label{tbl_ss_fam}
\end{center}
\end{table*}

\begin{table*}[h!]
  \begin{center}
    \small
\begin{tabular}{ll|rrrl}											\toprule
Avatar	&	Text	&	Coef.	&	Std. Err.	&	p-value	&		\\	\midrule
Second Life	&	Facebook	&	-0.2289	&	0.0517	&	0.0001	&	***	\\	
	&	X	&	-0.0144	&	-0.0533	&	0.9998	&		\\	
	&	Instagram	&	-0.2430	&	0.0539	&	$<$0.001	&	***	\\	\midrule
ZEPETO	&	Facebook	&	-0.2541	&	0.0515	&	$<$0.001	&	***	\\	
	&	X	&	-0.0109	&	0.0530	&	0.9999	&		\\	
	&	Instagram	&	-0.2683	&	0.0534	&	$<$0.001	&	***	\\	\midrule
Pigg Party	&	Facebook	&	-0.2339	&	0.0513	&	$<$0.001	&	***	\\	
	&	X	&	-0.0093	&	-0.0528	&	1.0000	&		\\	
	&	Instagram	&	-0.2481	&	0.0532	&	$<$0.001	&	***	\\	\bottomrule
\end{tabular}
  \caption{
  Result of Tukey's range test in ANOVA for perceived social support from offline friends with controlling age, gender, and usage frequency, where $F(5, 8937) = 16.391; p\mathrm{-value} < 0.001$.
}
    \label{tbl_ss_fri}
\end{center}
\end{table*}

\begin{table*}[h!]
  \begin{center}
\small
\begin{tabular}{ll|rrrl}											\toprule
Avatar	&	Text	&	Coef.	&	Std. Err.	&	p-value	&		\\	\midrule
Second Life	&	Facebook	&	0.1006	&	0.0177	&	$<$0.001	&	***	\\	
	&	X	&	0.0189	&	0.0183	&	0.9071	&		\\	
	&	Instagram	&	0.0657	&	0.0185	&	0.0050	&	**	\\	\midrule
ZEPETO	&	Facebook	&	0.1001	&	0.0177	&	$<$0.001	&	***	\\	
	&	X	&	0.0184	&	0.0182	&	0.9131	&		\\	
	&	Instagram	&	0.0652	&	0.0183	&	0.0048	&	**	\\	\midrule
Pigg Party	&	Facebook	&	0.1241	&	0.0176	&	$<$0.001	&	***	\\	
	&	X	&	0.0424	&	0.0181	&	0.1770	&		\\	
	&	Instagram	&	0.0892	&	0.0182	&	$<$0.001	&	***	\\	\bottomrule
\end{tabular}
  \caption{Result of Tukey's range test in ANOVA for loneliness with controlling age, gender, and usage frequency, where $F(5, 8937) =  19.299; p\mathrm{-value} < 0.001$.
}
    \label{tbl_Loneliness}
\end{center}
\end{table*}

\begin{table*}[h!]
  \begin{center}
\small
\begin{tabular}{ll|rrrl}											\toprule
Avatar	&	Text	&	Coef.	&	Std. Err.	&	p-value	&		\\	\midrule
Second Life	&	Facebook	&	0.8123	&	0.0298	&	$<$0.001	&	***	\\	
	&	X	&	0.9191	&	0.0307	&	$<$0.001	&	***	\\	
	&	Instagram	&	0.8920	&	0.0311	&	$<$0.001	&	***	\\	\midrule
ZEPETO	&	Facebook	&	0.7407	&	0.0297	&	$<$0.001	&	***	\\	
	&	X	&	0.8476	&	0.0305	&	$<$0.001	&	***	\\	
	&	Instagram	&	0.8204	&	0.0308	&	$<$0.001	&	***	\\	\midrule
Pigg Party	&	Facebook	&	0.6726	&	0.0296	&	$<$0.001	&	***	\\	
	&	X	&	0.7794	&	0.0304	&	$<$0.001	&	***	\\	
	&	Instagram	&	0.7523	&	0.0306	&	$<$0.001	&	***	\\	\bottomrule
\end{tabular}
  \caption{Result of Tukey's range test in ANOVA for bullying victimization with controlling age, gender, and usage frequency, where $F(5, 8937) = 465.833; p\mathrm{-value} < 0.001$.
    }
    \label{tbl_bullying}
\end{center}
\end{table*}

\begin{table*}[h!]
  \begin{center}
\small
\begin{tabular}{ll|rrl}	\toprule
Avatar	&	Text	&	Diff.	&	p-value	&		\\\midrule
Second Life	&	Facebook	&	0.5093	&	$<$0.001	&	***	\\
	&	X	&	0.5108	&	$<$0.001	&	***	\\
	&	Instagram	&	0.5734	&	$<$0.001	&	***	\\\midrule
ZEPETO	&	Facebook	&	0.3822	&	$<$0.001	&	***	\\
	&	X	&	0.3838	&	$<$0.001	&	***	\\
	&	Instagram	&	0.4464	&	$<$0.001	&	***	\\\midrule
Pigg Party	&	Facebook	&	0.4164	&	$<$0.001	&	***	\\
	&	X	&	0.4179	&	$<$0.001	&	***	\\
	&	Instagram	&	0.4805	&	$<$0.001	&	***	\\
 	\bottomrule
\end{tabular}
  \caption{
  Comparison of the regression coefficients of social support from family in avatar and text communication services using Student t-test with controlling age, gender, and usage frequency
    }
    \label{tbl_oss_famss}
\end{center}
\end{table*}

\begin{table*}[h!]
  \begin{center}
\small
\begin{tabular}{ll|rrrl}	\toprule
Avatar	&	Text	&	Diff.	&	p-value	&		\\\midrule
Second Life	&	Facebook	&	0.3475	&	$<$0.001	&	***	\\
	&	X	&	0.3462	&	$<$0.001	&	***	\\
	&	Instagram	&	0.4560	&	$<$0.001	&	***	\\\midrule
ZEPETO	&	Facebook	&	0.2469	&	$<$0.001	&	***	\\
	&	X	&	0.2456	&	$<$0.001	&	***	\\
	&	Instagram	&	0.3554	&	$<$0.001	&	***	\\\midrule
Pigg Party	&	Facebook	&	0.2923	&	$<$0.001	&	***	\\
	&	X	&	0.2910	&	$<$0.001	&	***	\\
	&	Instagram	&	0.4009	&	$<$0.001	&	***	\\	\bottomrule
\end{tabular}
  \caption{
  Comparison of the regression coefficients of social support from friends in avatar and text communication services using Student t-test with controlling age, gender, and usage frequency
    }
    \label{tbl_oss_friss}
\end{center}
\end{table*}

\newpage

\begin{figure*}[t!]
  \begin{center}
    \includegraphics[width=1.7\columnwidth]{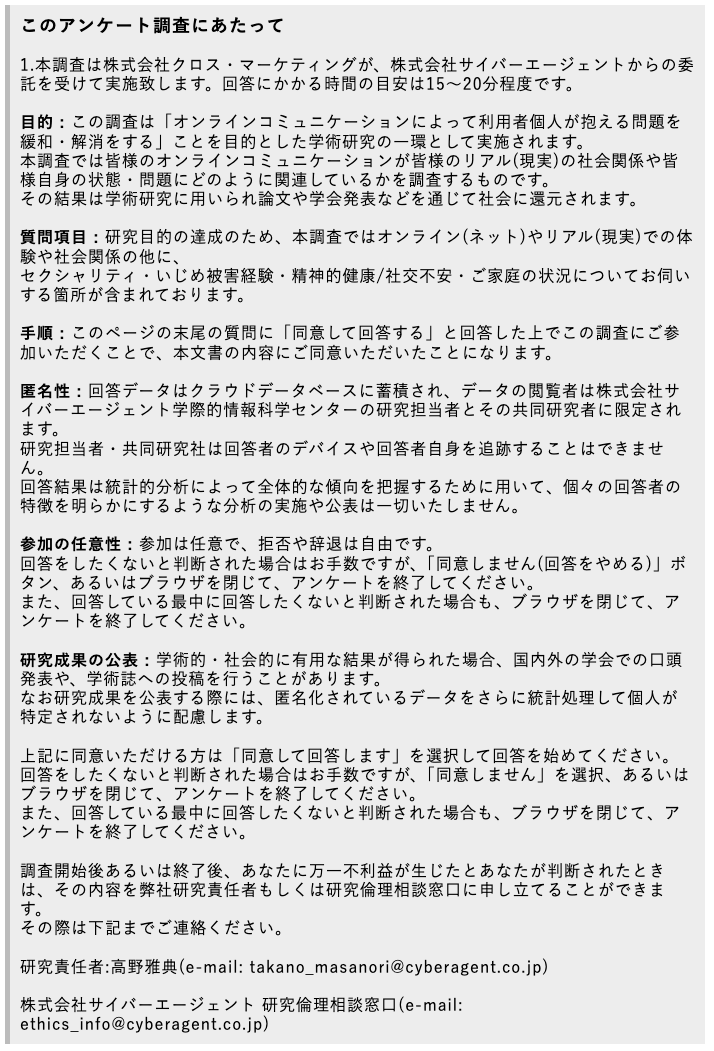}
    \caption{The screenshot of the informed consent on this research survey (original Japanese version)..
    See also Fig.~\ref{fig_ss_informed_consent_en} for the English version.
        }
    \label{fig_ss_informed_consent}
  \end{center}
\end{figure*}

\begin{figure*}[t!]
  \begin{center}
    \includegraphics[width=1.7\columnwidth]{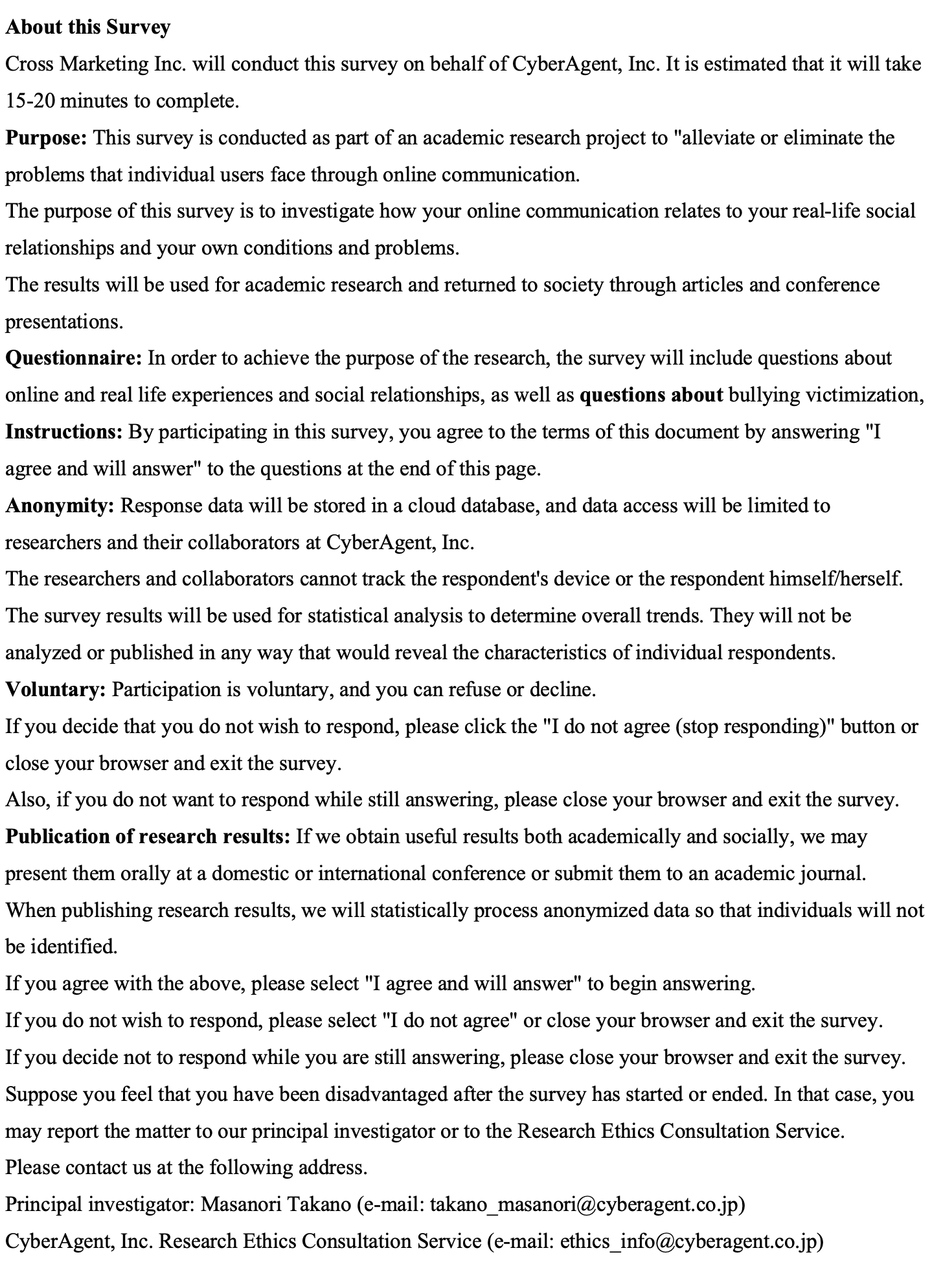}
    \caption{English version of the screenshot of the informed consent on this research survey..
    We did not use this version because all survey were conducted in Japanese.
        }
    \label{fig_ss_informed_consent_en}
  \end{center}
\end{figure*}

\end{document}